\documentstyle[epsf,preprint]{ptptex}
\begin{document}
\begin{flushright}
CU-TP/99-01
\end{flushright}
\begin{center}
{\large \bf Atmospheric Neutrino Oscillations in Three-Flavor\\
 Neutrinos. II}
\vspace{0.2cm}\\
Toyokazu {\sc Sakai} and Tadayuki {\sc Teshima}\footnote{E-mail: teshima@isc.chubu.ac.jp}\\
{\it Department of Applied Physics,  Chubu University, Kasugai 487-8501, Japan}
\vspace{0.3cm}\\
\begin{minipage}[t]{15cm}
{\small \setlength{\baselineskip}{14pt}
\par
 We analyze the atmospheric neutrino experiments of SuperKamiokande (535 days) using the three-flavor neutrino framework with the mass hierarchy $m_1\approx m_2\ll m_3$. We study the event ratios of sub-GeV, multi-GeV and upward through-going muons zenith angle distributions. Taking into account the atmospheric and terrestrial experiments, we obtain the allowed regions of mass and mixing parameters $\Delta m_{12}^2,\ \sin^22\theta_{12},\ \Delta m_{23}^2,\ \theta_{13}$ and $\theta_{23}$. The mass parameter $\Delta m^2_{23}$ is restricted to 0.01\,{eV$^2$}-0.0002\,{eV$^2$}. For mixing parameters, there is no difference between the large $\theta_{12}$ angle solution and the small one, and $\theta_{13}<13^\circ,\ 28^\circ<\theta_{23}<37^\circ$ and $53^\circ<\theta_{23}<62^\circ$ for $\Delta m_{23}^2=0.01\,{\rm eV}^2$, $\theta_{13}<23^\circ$ and $29^\circ<\theta_{23}<61^\circ$ for $\Delta m_{23}^2=0.002\,{\rm eV}^2$, $4^\circ<\theta_{13}<22^\circ$ and $38^\circ<\theta_{23}<54^\circ$ for $\Delta m_{23}^2=0.0002\,{\rm eV}^2$.
 
 }
\end{minipage}
\end{center}
\vspace{0.3cm}
%%%%%%%%%%%%% section 1%%%%%%%%%%%%%%%
\section{Introduction}
 \setlength{\baselineskip}{16pt}
SuperKamiokande experiment \cite{SUPERKAMIOKANDE} has observed a definite atmospheric neutrino anomaly and confirmed a hypothesis of neutrino oscillation. In their two-flavor mixing analyses of the sub-GeV and multi-GeV zenith angle distribution, it has been obtained that the $\nu_\mu\leftrightarrow\nu_\tau$ oscillation is preferred over the $\nu_\mu\leftrightarrow\nu_e$ oscillation and the range of mass parameter $\Delta m^2$ is from $3\times10^{-4}{\rm eV}^2$ to $0.8\times10^{-2}{\rm eV}^2$.\cite{SUPERKAMIOKANDE} 
\par
Observing the neutrino oscillations is one of the most important experiments suggesting a new physics beyond the standard model. Constructing a model beyond the standard model, it is necessary to confirm the mixing and mass parameters of three neutrinos in the low energy region although there is no restriction on number of neutrinos in very high energy region.\cite{MOHAPATRA} The three-flavor neutrinos scenario\cite{THREE} is found to be consistent with above the large $\nu_\mu\leftrightarrow\nu_\tau$ oscillation and the solar neutrino anomaly\cite{SAGE} and terrestrial neutrino experiments.\cite{E531,E776,CDHSW}  K2K, MINOS and CERN-Gran Sasso long-baseline experiments \cite{K2K} are in preparation to confirm the $\nu_\mu\leftrightarrow\nu_e$ and $\nu_\mu\leftrightarrow\nu_\tau$ oscillation precisely. Then it becomes important to examine the mass and mixing parameters in the three-flavor neutrinos framework\cite{THREE,FOGLIA} precisely. 
\par 
In this paper, we analyze the SuperKamiokande atmospheric neutrino experiment (535 days) of the sub-GeV, multi-GeV and upward through-going muons zenith angle distributions in the three-flavor neutrino framework with a hierarchy of neutrino masses $m_1\approx m_2\ll m_3$. In a previous work,\cite{TESHIMA} we have analyzed the zenith angle distributions in the SuperKamiokande atmospheric neutrino experiment(325.8 days).\cite{SUPERKAMIOKANDEP} In that work, we analyzed the double ratios of the sub-GeV and multi-GeV zenith angle distributions neglecting the matter effects of the Earth and the difference between incident neutrinos and detected charged leptons. 
\par
In the three-flavor neutrino framework with the hierarchy $m_1\approx m_2\ll m_3$, there are 5 parameters $\theta_{12},\ \theta_{13},\ \theta_{23},\ \Delta m^2_{12}=m_2^2-m_1^2,\ \Delta m^2_{23}=m_3^2-m_2^2 $ concerned with the neutrino oscillation. For solar neutrino deficit, the MSW solution\cite{KUO,FOGLIS,TESHIMA} taking into account the matter effects predicts the large angle solution $\sin^22\theta_{12}=0.6$--$0.9$, $\Delta m^2_{12}=4\times10^{-6}$--$7\times10^{-5}{\rm eV}^2$ and small angle solution $\sin^22\theta_{12}=0.003$--$0.01$, $\Delta m^2_{12}=3\times10^{-6}$--$1.2\times10^{-5}{\rm eV}^2$ for $\theta_{13}=0^\circ$--$20^\circ$, and these large and small angle solution are merged for $\theta_{13}=25^\circ$--$50^\circ$. The vacuum solution for solar neutrinos is obtained as $\Delta m^2_{12}\sim10^{-10}{\rm eV}^2$.\cite{VACUUME} \ In terrestrial neutrino experiments, Fogli {\it et al.}\cite{FOGLIT} have thoroughly analyzed and obtained the allowed regions in $\tan^2\theta_{13}$--$\tan^2\theta_{23}$ plane for various values of the mass parameter $\Delta m^2_{23}$. We also analyzed the terrestrial neutrino experiments,\cite{TESHIMA} and obtained results similar to those of Fogli et al.  The allowed region in $\tan^2\theta_{13}-\tan^2\theta_{23}$ plane was more restricted for higher values of $\Delta m^2_{23}$, and  the allowed region spread out for lower values of it. In this paper,  we combine the results obtained from the atmospheric neutrino experiments of SuperKamiokande and the terrestrial experiments to obtain the allowed regions for mixing and mass parameters in three-flavor neutrinos. 

%%%%%%%%%%% section 2 %%%%%%%%%%%%%%
\section{Neutrino oscillation in three-flavor neutrinos}
\par 
The unitary matrix $U$ representing the neutrino mixing is defined as 
\begin{equation}
\nu_{l_\alpha}=\sum^3_{\beta=1} U_{l_\alpha\beta}\nu_\beta,\ \ \ l_\alpha=e,\ \mu, \tau, 
\end{equation}
where the states $\nu_{l_\alpha}$ and $\nu_\beta$ are the flavor and mass eigenstate of neutrinos, respectively. 
This mixing matrix corresponds to the CKM matrix $U^\dagger_{\rm CKM}$ for quarks sector. We parameterize the unitary matrix neglecting the {\it CP} violation phases as 
\begin{eqnarray}
 U&=&\exp{(i\theta_{23}\lambda_7)}\exp{(i\theta_{13}\lambda_5)}\exp{(i\theta_{12}
             \lambda_2)}\nonumber\\
            &=&\left(
      \begin{array}{ccc}
      c_{12}^{\nu}c_{13}^{\nu} & s_{12}^{\nu}c_{13}^{\nu} & s_{13}^{\nu} \\
      -s_{12}^{\nu}c_{23}^{\nu}-c_{12}^{\nu}s_{23}^{\nu}s_{13}^{\nu} & c_{12}^{\nu}c_{23}^{\nu}-s_{12}^{\nu}s_{23}^{\nu}s_{13}^{\nu} & s_{23}^{\nu}c_{13}^{\nu} \\
      s_{12}^{\nu}s_{23}^{\nu}-c_{12}^{\nu}c_{23}^{\nu}s_{13}^{\nu} & -c_{12}^{\nu}s_{23}^{\nu}-s_{12}^{\nu}c_{23}^{\nu}s_{13}^{\nu} & c_{23}^{\nu}c_{13}^{\nu} 
      \end{array}\right), \\
 & & \quad\quad c_{ij}^{\nu}=\cos{\theta}^{\nu}_{ij},\ \  s_{ij}^{\nu}=\sin{\theta}^{\nu}_{ij}, \nonumber 
\end{eqnarray}  
where the $\lambda_i$ are the Gell-Mann matrices.
\par 
The probabilities for transitions  $\nu_{l_\alpha} \to \nu_{l_\beta}$ are written as 
\begin{eqnarray}
P(\nu_{l_\alpha}\to\nu_{l_\beta})&=&|<\nu_{l_\beta}(t)|\nu_{l_\alpha}(0)>|^2 = \delta_{l_\alpha l_\beta}+p_{\nu_{l_\alpha}\to\nu_{l_\beta}}^{12}S_{12}\nonumber\\
&&\qquad+p_{\nu_{l_\alpha}\to\nu_{l_\beta}}^{23}S_{23}+p_{\nu_{l_\alpha}\to\nu_{l_\beta}}^{31}S_{31},\nonumber \\
p_{\nu_{l_\alpha}\to\nu_{l_\beta}}^{12}&=&-2\delta_{l_\alpha l_\beta}(1-2U_{l_\alpha3}^2)+2(U_{l_\alpha1}^2U_{l_\beta1}^2+U_{l_\alpha2}^2U_{l_\beta2}^2-U_{l_\alpha3}^2U_{l_\beta3}^2), \nonumber \\       
p_{\nu_{l_\alpha}\to\nu_{l_\beta}}^{23}&=&-2\delta_{l_\alpha l_\beta}(1-2U_{l_\alpha1}^2)+2(-U_{l_\alpha1}^2U_{l_\beta1}^2+U_{l_\alpha2}^2U_{l_\beta2}^2+U_{l_\alpha3}^2U_{l_\beta3}^2), \nonumber \\
p_{\nu_{l_\alpha}\to\nu_{l_\beta}}^{31}&=&-2\delta_{l_\alpha l_\beta}(1-2U_{l_\alpha2}^2)+2(U_{l_\alpha1}^2U_{l_\beta1}^2-U_{l_\alpha2}^2U_{l_\beta2}^2+U_{l_\alpha3}^2U_{l_\beta3}^2) ,\nonumber \\
&&
\end{eqnarray}
where $S_{ij}$ is the term representing the neutrino oscillation;  
\begin{equation}
S_{ij}=\sin^21.27\frac{\Delta m^2_{ij}}{E}L.
\end{equation}
Here $\Delta m^2_{ij}=|m^2_i-m^2_j|$, $E$ and $L$  are measured in units eV$^2$, GeV and km, respectively.
\par
The values of neutrino masses are not known precisely, but two mass parameters are necessary to account for the solar neutrino experiments and the atmospheric neutrino experiments. In the former, mass parameter $\Delta m^2$ is obtained as $10^{-4}$--$10^{-5}\,{\rm eV}^2$\cite{SAGE,FOGLIS,TESHIMA} or $\sim 10^{-10}{\rm eV}^2$,\cite{VACUUME} and in the latter, $\Delta m^2$ is obtained as $10^{-1} - 10^{-3}{\rm eV}^2$.\cite{SUPERKAMIOKANDEP,SUPERKAMIOKANDE2} \ Then it seems that the most reasonable mass hierarchy is that in which the lower two neutrino masses of three neutrinos are very close and the third is rather far from them. Thus we assume that three neutrino masses have a mass hierarchy obeying  
\begin{equation}
m_1 \approx m_2 \ll m_3. \label{masshie}
\end{equation}
With the hierarchy Eq.~(\ref{masshie}), $\Delta m_{12}^2\ll\Delta m_{23}^2 \simeq \Delta m_{13}^2$, Eq. (3) for the transition probabilities $P(\nu_{l_{\alpha}}\to\nu_{l_{\beta}})$ can be rewritten as
\begin{eqnarray}
P(\nu_{l_\alpha}\to\nu_{l_\alpha})&=&1-2(1-2U_{l_\alpha3}^2-U_{l_\alpha1}^4-U_{l_\alpha2}^4+U_{l_\alpha3}^4)S_{12}-4U_{l_\alpha3}^2(1-U_{l_\alpha3}^2)S_{23}, \nonumber \\
P(\nu_{l_\alpha}\to\nu_{l_\beta})&=&P(\nu_{l_\beta}\to\nu_{l_\alpha})=2(U_{l_\alpha1}^2U_{l_\beta1}^2+U_{l_\alpha2}^2U_{l_\beta2}^2-U_{l_\alpha3}^2U_{l_\beta3}^2)S_{12}  \nonumber\\
&&\qquad\qquad\qquad+4U_{l_\alpha3}^2U_{l_\beta3}^2S_{23}.  
\label{trans.prob.}
\end{eqnarray}
\par
In the atmospheric neutrino experiments, especially in the sub-GeV ones, the matters of the Earth have the important effects. Matter effects are induced by the quantity $A=2\sqrt{2}EG_FN_e$. In the Earth, $N_e\sim3 {\rm mol/cm^3}$, thus $A\sin2\theta_{13}<<1$ and we can approximate the mixing matrix $U$ as\cite{KUO}
\begin{equation}
U^M=\exp(i\lambda_7\theta_{23})\exp(i\lambda_5\theta_{13})\exp(i\lambda_2\theta^M_{12}),
\end{equation}
where
$$
\begin{array}{l}
\sin2\theta^M_{12}=\frac{\Delta m^2_{12}}{\Delta m^{M2}_{12}}\sin2\theta_{12},\\
\Delta m^{M2}_{12}=m^{M2}_2-m^{M2}_1,\ \ \Sigma=m^2_1+m^2_2,\\
m^{M2}_{1,2}=\frac12\left\{(\Sigma+A\cos^2\theta_{13})\mp\sqrt{(A\cos^2\theta_{13}-\Delta m^2_{12}\cos2\theta_{12})^2+(\Delta m^2_{12}\sin2\theta_{12})^2}\right\},\\
m^{M2}_3=m^2_3+A\sin^2\theta_{13}.\\
\end{array}
$$
This expression corresponds to the neutrino case, and  $A$ enters in this expression with opposite sign for the anti-neutrino case. If we approximate the density of the Earth to be uniform, we can take the expressions for the transition probability with matter effects as 
\begin{eqnarray}
P^M(\nu_{l_\alpha}\to\nu_{l_\alpha})&=&1-2(1-2U^{M2}_{l_\alpha3}-U^{M4}_{l_\alpha1}-U^{M4}_{l_\alpha2}+U^{M4}_{l_\alpha3})S^M_{12}-4U^{M2}_{l_\alpha3}(1-U^{M2}_{l_\alpha3})S^M_{23},\nonumber\\
P^M(\nu_{l_\alpha}\to\nu_{l_\beta})&=&P^M(\nu_\beta\to\nu_\alpha)=2(U^{M2}_{l_\alpha1}U^{M2}_{l_\beta1}+U^{M2}_{l_\alpha2}U^{M2}_{l_\beta2}-U^{M2}_{l_\alpha3}U^{M2}_{l_\beta3})S^M_{12}\nonumber\\
&&\qquad +4U^{M2}_{l_\alpha3}U^{M2}_{l_\beta3}S^M_{23},
\end{eqnarray}
where 
\begin{equation}
S^M_{ij}=\sin^21.27\frac{\Delta m^{M2}_{ij}}{E}L.
\end{equation} 
%%%%%%%%% section 3%%%%%%%%%%%%%%%%%%%%%%%%
\section{Numerical analyses of neutrino oscillations}
\subsection{Atmospheric neutrinos}
Evidence for an anomaly in atmospheric neutrino experiments was pointed out in reports of Kamiokande collaboration \cite{HIRATA,FUKUDA} and IMB collaboration \cite{IMB} using water-Cherencov experiments. More recently, reports of SuperKamiokande collaboration \cite{SUPERKAMIOKANDEP,SUPERKAMIOKANDE} have provided more precise results on the anomaly in atmospheric neutrinos. In our previous papers,\cite{TESHIMA} we have analyzed the double ratios  $R(\mu/e)\equiv{R_{\rm expt}(\mu/e)}/{R_{\rm MC}(\mu/e)}$ of atmospheric neutrinos. In this work, we analyze the event ratio $N_{\rm Exp}(l_\alpha)/N_{\rm MC}(l_\alpha)$, where $l_\alpha$ represents $\mu$ and $e$. 

%%%%%%%%%%%%%%%%%%%%%%%%%%%%%%%%%%%%%%%%%%%%%
\begin{table}
\begin{center}
\caption{e-like, $\mu$-like atmospheric neutrino data and upward through-going muons fluxes of SuperKamiokande experiments (535days data). These values are the ratios of experimental data and Monte-Carlo simulations which  are obtained from the graphs in Ref.~1). $\mu$-like events include fully contained and partially contained events, and errors represent statistical ones only.}
\label{table1}
\vspace*{0.2cm}
Sub-GeV data
\begin{tabular}{ccc}\hline\hline
$\cos\theta$ range & $e$-like event ratio $N_{\rm Exp}/N_{\rm MC}$ & $\mu$-like event ratio $N_{\rm Exp}/N_{\rm MC}$ \\ \hline
-1.0 $-$ -0.6  & $1.37\pm0.09$ & $0.56\pm0.04$ \\
-0.6 $-$ -0.2  & $1.12\pm0.08$ & $0.71\pm0.05$ \\
-0.2 $-$ 0.2 & $1.18\pm0.08$ & $0.74\pm0.05$ \\
0.2 $-$ 0.6 & $1.05\pm0.08$ & $0.86\pm0.06$ \\
0.6 $-$ 1.0 & $1.15\pm0.08$ & $0.82\pm0.05$ \\ \hline
\end{tabular}
\vspace*{0.5cm}\\
Multi-GeV data
\begin{tabular}{ccc}\hline\hline
$\cos\theta$ range & $e$-like event ratio $N_{\rm Exp}/N_{\rm MC}$ & $\mu$-like event ratio $N_{\rm Exp}/N_{\rm MC}$ \\ \hline
-1.0 $-$ -0.6  & $1.35\pm0.21$ & $0.56\pm0.07$ \\
-0.6 $-$ -0.2  & $1.10\pm0.16$ & $0.57\pm0.07$ \\
-0.2 $-$ 0.2 & $1.13\pm0.15$ & $0.79\pm0.07$ \\
0.2 $-$ 0.6 & $1.42\pm0.18$ & $1.02\pm0.09$ \\
0.6 $-$ 1.0 & $1.18\pm0.20$ & $1.03\pm0.10$ \\ \hline
\end{tabular}
\vspace*{0.5cm}\\
Upward through-going $\mu$ data
\begin{tabular}{cc}\hline\hline
$\cos\theta$ range & $\mu$ flux ratio $N_{\rm Exp}/N_{\rm MC}$  \\ \hline
-1.0 $-$ -0.9 & $0.77\pm0.14$  \\
-0.9 $-$ -0.8 & $0.79\pm0.12$  \\
-0.8 $-$ -0.7 & $0.58\pm0.11$  \\
-0.7 $-$ -0.6 & $0.96\pm0.13$  \\
-0.6 $-$ -0.5 & $0.73\pm0.10$  \\
-0.5 $-$ -0.4 & $0.80\pm0.10$  \\
-0.4 $-$ -0.3 & $0.74\pm0.10$  \\
-0.3 $-$ -0.2 & $0.92\pm0.10$  \\
-0.2 $-$ -0.1 & $1.02\pm0.10$  \\
-0.1 $-$  0.0 & $1.09\pm0.10$  \\ \hline
\end{tabular}
\end{center}
\end{table}
%%%%%%%%%%%%%%%%%%%%%%%%%%%%%%%%%%%%%%%%%%%%%
\par
The zenith angle distributions of sub-GeV, multi-GeV events and upward through-going $\mu$ fluxes are taken from the SuperKamiokande 535 days experiments.\cite{SUPERKAMIOKANDE} The data are tabulated in Table I. These values are taken from the experimental event data and Monte-Carlo simulations which are given graphically in Ref. 1).  $\mu$-like events include the fully contained and partially contained events. Errors represent statistical ones only.
In the above data, sub-GeV experiments detect the visible-energy less than 1.33GeV. 
\par
The zenith angle $\theta$ dependent events ${dN_{\rm Exp}(l_\alpha)/d\cos\theta}$ and ${dN_{\rm MC}(l_\alpha)/d\cos\theta}$ are defined as
\begin{subequations}
\begin{equation}
{\frac{dN_{\rm Exp}(l_\alpha)}{d\cos\theta}}=\sum_\beta\int\epsilon_{\alpha}(E_{\alpha})\sigma_{\alpha}(E_{\nu},E_{\alpha},\psi)F_{\beta}(E_{\nu},\theta-\psi)P^M(\nu_\beta\to\nu_\alpha)dE_{\alpha}dE_{\nu}d\cos\psi,
\end{equation}
\begin{equation}
{\frac{dN_{\rm MC}(l_\alpha)}{d\cos\theta}}=\int\epsilon_{\alpha}(E_{\alpha})\sigma_{\alpha}(E_{\nu},E_{\alpha},\psi)F_{\alpha}(E_{\nu},\theta-\psi)dE_{\alpha}dE_{\nu}d\cos\psi,
\end{equation}
\end{subequations}
where the summation $\sum_\beta$ are taken over $\nu_\mu$ and $\nu_e$. In these expressions, processes of $\bar{\nu}_\mu$ and$\bar{\nu}_e$ are contained. $\epsilon_{\alpha}(E_{\alpha})$ is the detection efficiency of the detector for $\alpha$-type charged lepton with energy $E_{\alpha}$. $\sigma_{\alpha}(E_\nu,E_\alpha,\psi)$ is the differential cross section of scattering $l_\alpha$ with energy $E_\nu$ by incident $\nu_{\alpha}$ with energy $E_\alpha$, where angle $\psi$ is the scattering angle relative to the direction of incident $\nu_\alpha$. $F_{\beta}(E_{\nu},\theta)$ is the incident $\nu_{\beta}$ flux with energy $E_{\nu}$ at the zenith angle $\theta$.   $P^M(\nu_\alpha\to\nu_\beta)$ is the transition probability with the matter effects expressed in Eq.~(8) and it depends on the energy $E_\nu$ and the distance $L$. This distance depends on the zenith angle $\theta$ as $L=\sqrt{(r+h)^2-r^2\sin^2\theta}-r\cos\theta$, where $r$ is the radius of the Earth and $h$ is the altitude of the production point of atmospheric neutrinos. 
\par
Informations of $F_\beta(E_\nu,\theta)$ for multi-GeV case are given in Refs.~\citen{GAISSER,HONDA,AGRAWAL}. Informations of $F_\beta(E_\nu,\theta)$ including the geomagnetic effects for sub-GeV case are taken from Ref.~\citen{HONDAP}. Other informations of $\epsilon_\alpha(E_\alpha)$ and $\sigma_\alpha(E_\nu,E_\alpha,\psi)$ are given in Ref.~\citen{KAJITA}. The upward through-going muons fluxes are simulated in Ref.~\citen{KAJITA}. Explicit calculation of Eq.~(10) is explained in Appendix A.

%%%%%%%%%%%%%%%%%%% fig.1%%%%%%%%%%%%%%%%%%%%%%
\begin{figure}
   \epsfysize=13cm
   \centerline{\epsfbox{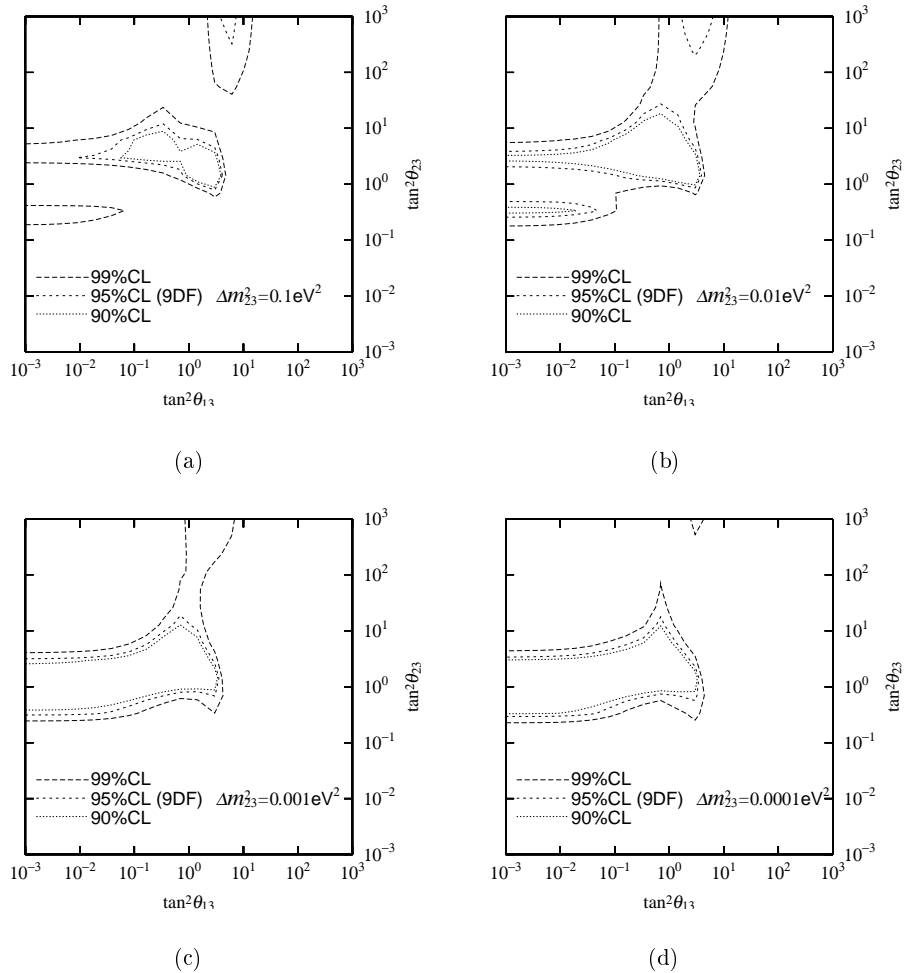}}
%  \figurebox{13cm}{13cm}
\caption{The plots of allowed regions in the $\tan^2\theta_{13}$--$\tan^2\theta_{23}$ plane determined by the zenith angle distributions of SuperKamiokande 535 days data. These figures correspond to the large $\theta_{12}$ angle solution, $\Delta m_{12}^2=3\times10^{-5}{\rm eV}^2$ and $\sin^22\theta= 0.7$. In these figures, broken thick, broken thin and dotted curves denote the regions allowed in 99\%, 95\% and 90\% C.L., respectively. Figs.~(a)--(d) show the plots of sub-GeV experiments.}
\end{figure}   
\setcounter{figure}{0}
\begin{figure}
   \epsfysize=13cm
   \centerline{\epsfbox{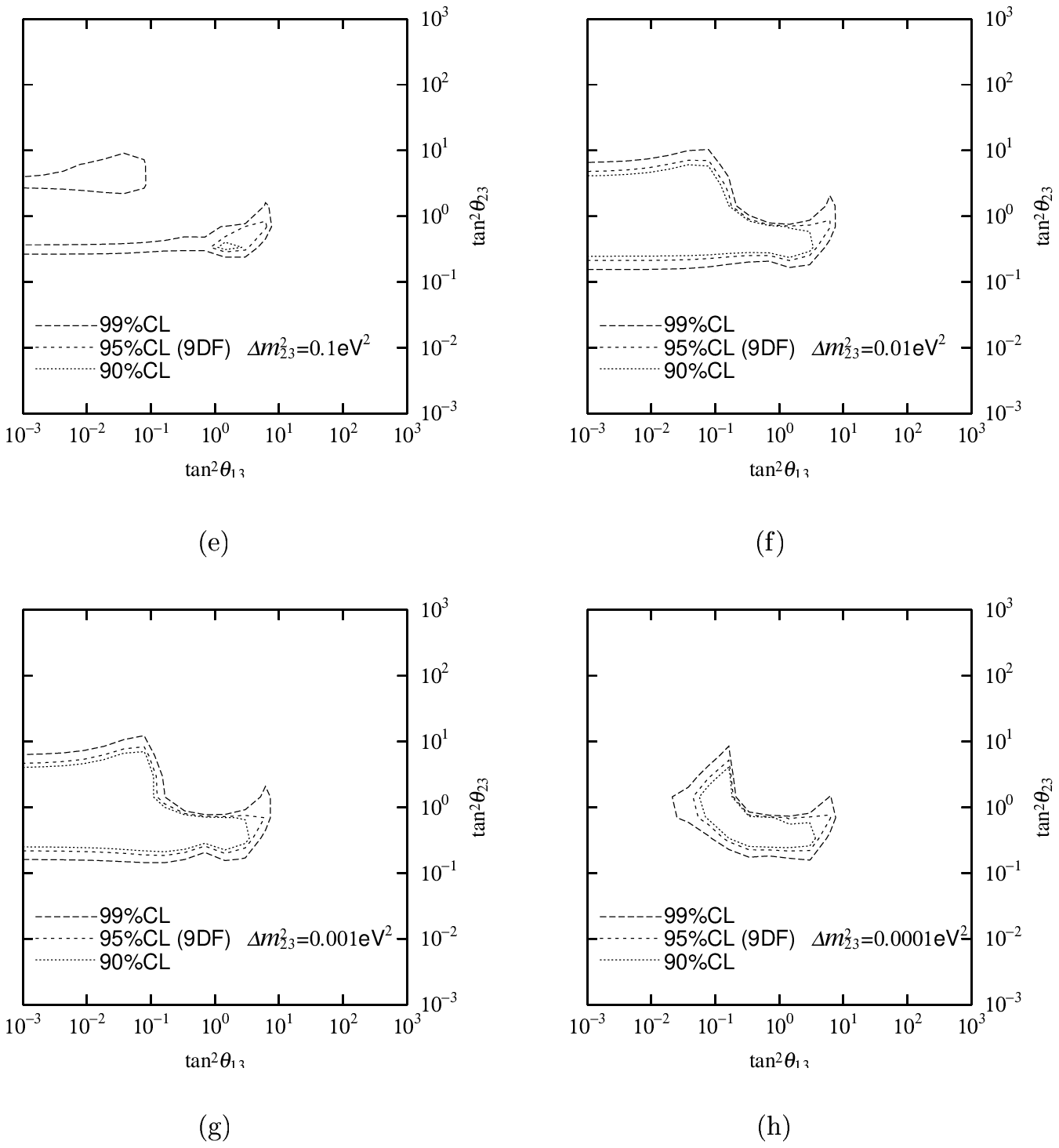}}
%  \figurebox{13cm}{13cm}
\caption{The plots of allowed regions in the $\tan^2\theta_{13}$--$\tan^2\theta_{23}$ plane.  Figs.~(e)--(h) the plots of multi-GeV experiments.}   
\end{figure}
\setcounter{figure}{0}   
\begin{figure}
   \epsfysize=13cm
   \centerline{\epsfbox{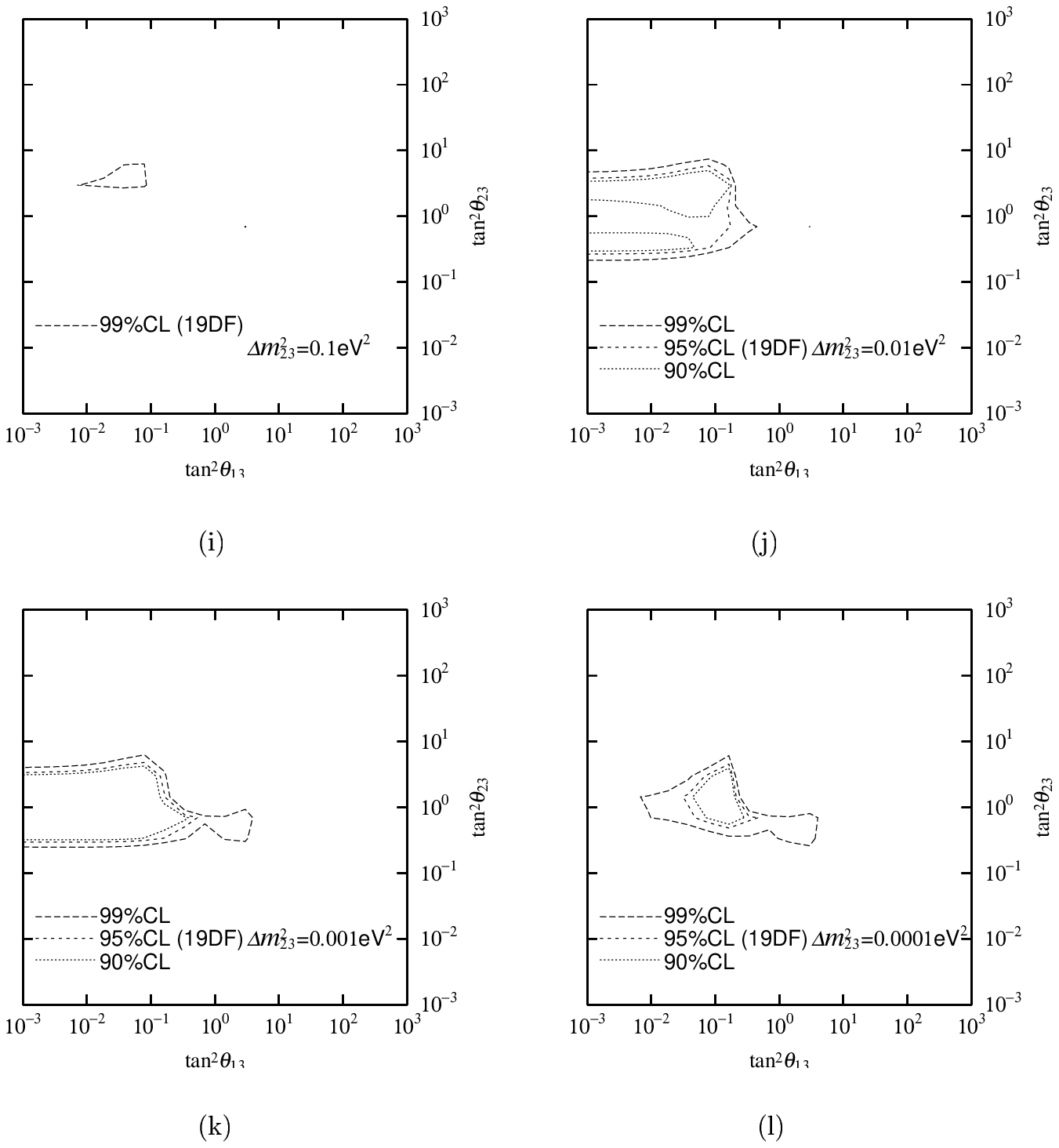}}
%  \figurebox{13cm}{13cm}  
\caption{The plots of allowed regions in the $\tan^2\theta_{13}$--$\tan^2\theta_{23}$ plane. Figs.~(i)--(l) show the plots of the combinations of the sub-GeV and multi-GeV experiments.}  
\label{fig1}
\end{figure}  
%%%%%%%%%%%%%%%%%%%%%%%%%%%%%%%%%%%%%%%%%%%%%%%

%%%%%%%%%%%%%%%%%%% fig.2%%%%%%%%%%%%%%%%%%%%%%
\begin{figure}
   \epsfysize=13cm
   \centerline{\epsfbox{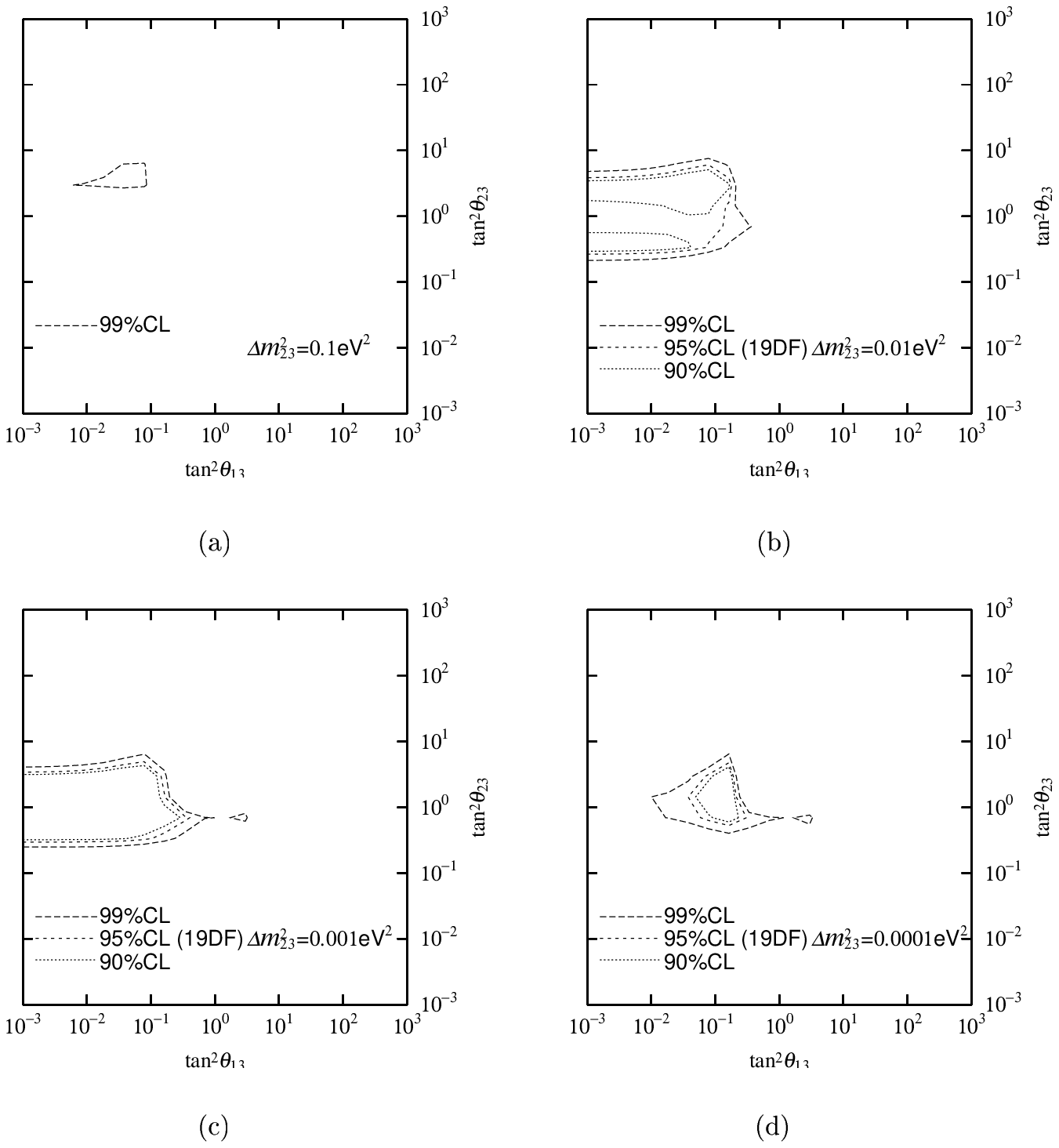}}
%  \figurebox{13cm}{13cm}
\caption{The plots of allowed regions on $\tan^2\theta_{13}$--$\tan^2\theta_{23}$ plane determined by the zenith angle distributions of the combination of sub-GeV and multi-GeV $\mu$ SuperKamiokande 535 days data. These figures correspond to the small $\theta_{12}$ angle solution, $\Delta m_{12}^2=10^{-5}{\rm eV}^2$ and $\sin^22\theta-0.005$. In these figures, broken thick, broken thin and dotted curves denote the regions allowed in 99\%, 95\% and 90\% C.L., respectively.}
\label{fig2} 
\end{figure}  
%%%%%%%%%%%%%%%%%%%%%%%%%%%%%%%%%%%%%%%%%%%%%%%

\par
Since $P^M(\nu_\alpha\to\nu_\beta)$ is the function of $\Delta m^2_{12}, \Delta m^2_{23}, \theta_{12}, \theta_{13}$ and $\theta_{23}$, the ratio ${(dN_{\rm Exp}(l_\alpha)}$    ${/d\cos\theta})/({dN_{\rm MC}(l_\alpha)/d\cos\theta})$ of the zenith angle distributions is a function of $\Delta m^2_{12}, \Delta m^2_{23}, \theta_{12},$  $ \theta_{13}, \theta_{23}$ and $\theta$. We analyze the atmospheric neutrino data fixing the parameters $\Delta m^2_{12}$ and $\sin^22\theta_{12}$, which have been determined from the solar neutrino experiments,\cite{SAGE,FOGLIS,TESHIMA} as $\Delta m_{12}^2=3\times10^{-5}{\rm eV}^2$ and $\sin^22\theta_{12}=0.7$, which corresponds to the large angle solution, and $\Delta m_{12}^2=10^{-5}{\rm eV}^2$ and $\sin^22\theta_{12}=0.005$, which corresponds to the small angle solution. We treat the ratios of the zenith angle distributions of the experimental events and the ones of Monte-Calro simulation, $(N_{\rm Exp}(l_\alpha)/N_{\rm MC}(l_\alpha))_i$, where $i$ represents the region number of the bins of zenith angle $\theta$. 
\par
We treat the  $\chi^2$  defined as
\begin{equation}
\chi^2=\sum_{i,\ l_\alpha}\frac{\left\{(N_{\rm Exp}(l_\alpha)/N_{\rm MC}(l_\alpha))^{\rm cal}_i-(N_{\rm Exp}(l_\alpha)/N_{\rm MC}(l_\alpha))^{\rm data}_i\right\}^2}{(\sigma_{st})^2_i+(\sigma_{sy})^2_i}.
\end{equation}
For the sub-GeV and multi-GeV experiments, the summation on $i$ are over 1 to 5 of zenith angle range bins and for the upward through-going $\mu$, 1 to 10. The summation on $l_\alpha$  are over $\mu$ and $e$ for the sub-GeV and multi-GeV experiments. We estimate the $\chi^2$ on the various of $\Delta m^2_{23}, \theta_{13}$ and $\theta_{23}$. $\sigma_{st}$ represents the statistical error and $\sigma_{sy}$ systematic one. We adopt the values of $\sigma_{sy}$ which are induced from the magnitudes of $10\%$  of the $N_{\rm MC}$ assumed as uncertainty. These values are deduced from the graphs given by the report of SuperKamiokande collaboration.\cite{SUPERKAMIOKANDE} In Fig. \ref{fig1}, we showed the contour plots of $\chi^2$ in the $\tan^2\theta_{13}$--$\tan^2\theta_{23}$ plane for various $\Delta m^2_{23}$. We showed the plots of sub-GeV experiment in Fig.~\ref{fig1}(a)--(d) and the plots of multi-GeV one in Fig.~\ref{fig1}(e)--(h) for the large $\theta_{12}$ angle solution,  $\Delta m_{12}^2=3\times10^{-5}{\rm eV}^2$ and $\sin^22\theta=0.7$. In these figures, broken thick, broken thin and dotted curves denote the regions allowed in 99\%, 95\% and 90\% C.L., respectively. The combinations of the sub-GeV and multi-GeV experiments are shown in Fig. \ref{fig1}~(i)--(l).  Fig.~\ref{fig2} represents the case of the small $\theta_{12}$ angle solution, $\Delta m^2_{12}=10^{-5}{\rm eV^2}$ and $\sin^22\theta=0.005$. 

\par
From these plots, we can get the following results for the mixing parameters:\\
\ \ (1) $\Delta m_{23}^2$ is allowed from $10^{-2}{\rm eV^2}$ to $2\times10^{-4}{\rm eV^2}$. The minimum $\chi^2$ is obtained as $14.5$ at $\Delta m^2_{23}=7\times 10^{-4}{\rm eV^2}$, $\tan^2\theta_{13}=6\times10^{-2}$ and $\tan^2\theta_{23}=2$.  The minimum $\chi^2$ in the restriction $\theta_{13}= 0$, which corresponds to the $\nu_\mu-\nu_\tau$ mixing, is $17$ at $\Delta m^2_{23}=2\times 10^{-3}{\rm eV^2}$. \\
\ \ (2) $\nu_e-\nu_\tau$ mixing is small and $\nu_\mu-\nu_\tau$ mixing is large; $\tan^2\theta_{13}<10^{-1}$ and $\tan^2\theta_{23}\sim 1$.\\
\ \  (3) There is no significant difference between the large $\theta_{12}$ solution and small one.\\
The result that $\Delta m_{23}^2$ is allowed from $10^{-2}{\rm eV^2}$ to $2\times10^{-4}{\rm eV^2}$ and the minimum $\chi^2$ in the restriction $\theta_{13}= 0$ is obtained at $\Delta m^2_{23}=2\times 10^{-3}{\rm eV^2}$ is the same as the result obtained by SuperKamiokande collabolation.\cite{SUPERKAMIOKANDE} 
\par
Plots of the upward through-going $\mu$ are shown in Fig.~\ref{fig3}(a)--(d) and the combinations of the sub-GeV, multi-GeV and upward through-going $\mu$ flux in Fig.~\ref{fig3}(e)--(h) for the large $\theta_{12}$ angle solution. The plots for the small $\theta_{12}$ angle solution is the similar to the Fig.~\ref{fig3}.  From these result,  we can say that\\ 
(1) the mixing parameters allowed in the sub-GeV and multi-GeV zenith angle distributions can explain the upward through-going $\mu$ flux experimental data,\\
(2) in large value of $\Delta m^2_{23}\sim10^{-1}$--$10^{-2}{\rm eV^2}$, the values near $45^\circ$ of $\theta_{23}$ are excluded, because the deficit from 1 of the event ratio $N_{\rm Exp}(\mu)/N_{\rm MC}(\mu)$ is not so large (about 0.2) in upward through-going $\mu$ flux compared with multi-GeV experiments. 

%%%%%%%%%%%%%%%%%%% fig.3%%%%%%%%%%%%%%%%%%%%%%
\begin{figure}
   \epsfysize=13cm
   \centerline{\epsfbox{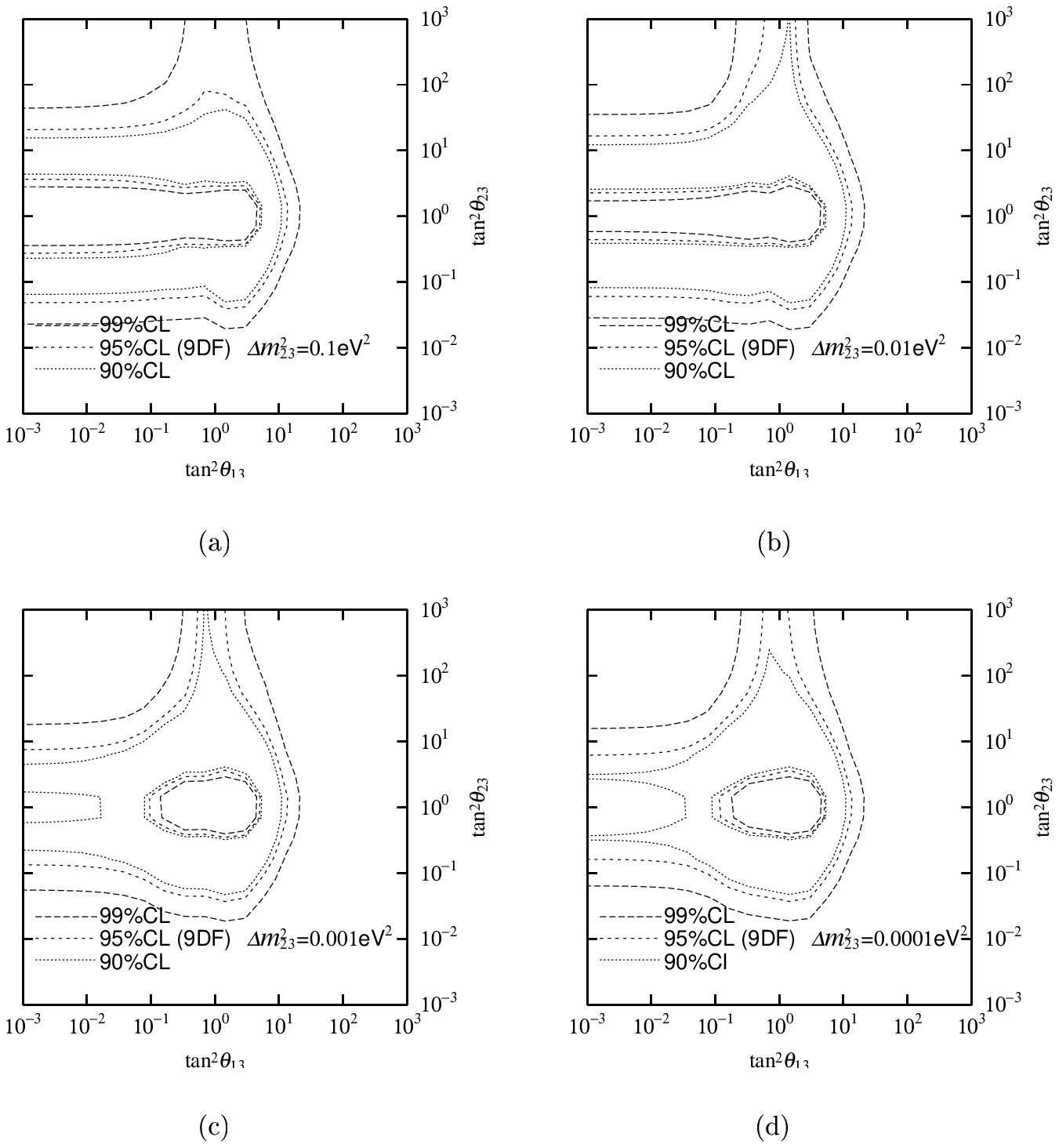}}
%  \figurebox{13cm}{13cm}
 \caption{The plots of allowed regions in the $\tan^2\theta_{13}$--$\tan^2\theta_{23}$ plane determined by the zenith angle distributions of the upward through-going $\mu$ SuperKamiokande data for large $\theta_{12}$ angle solution. There is no difference between the large $\theta_{12}$ angle solution and small one. In these figures, the broken thick, broken thin and dotted curves denote the regions allowed  in 99\%, 95\% and 90\% C.L., respectively. Fig.~(a)--(d) show the plot of the upward through-going $\mu$ experiments.}
\end{figure}  
\setcounter{figure}{2}
\begin{figure}
   \epsfysize=13cm
   \centerline{\epsfbox{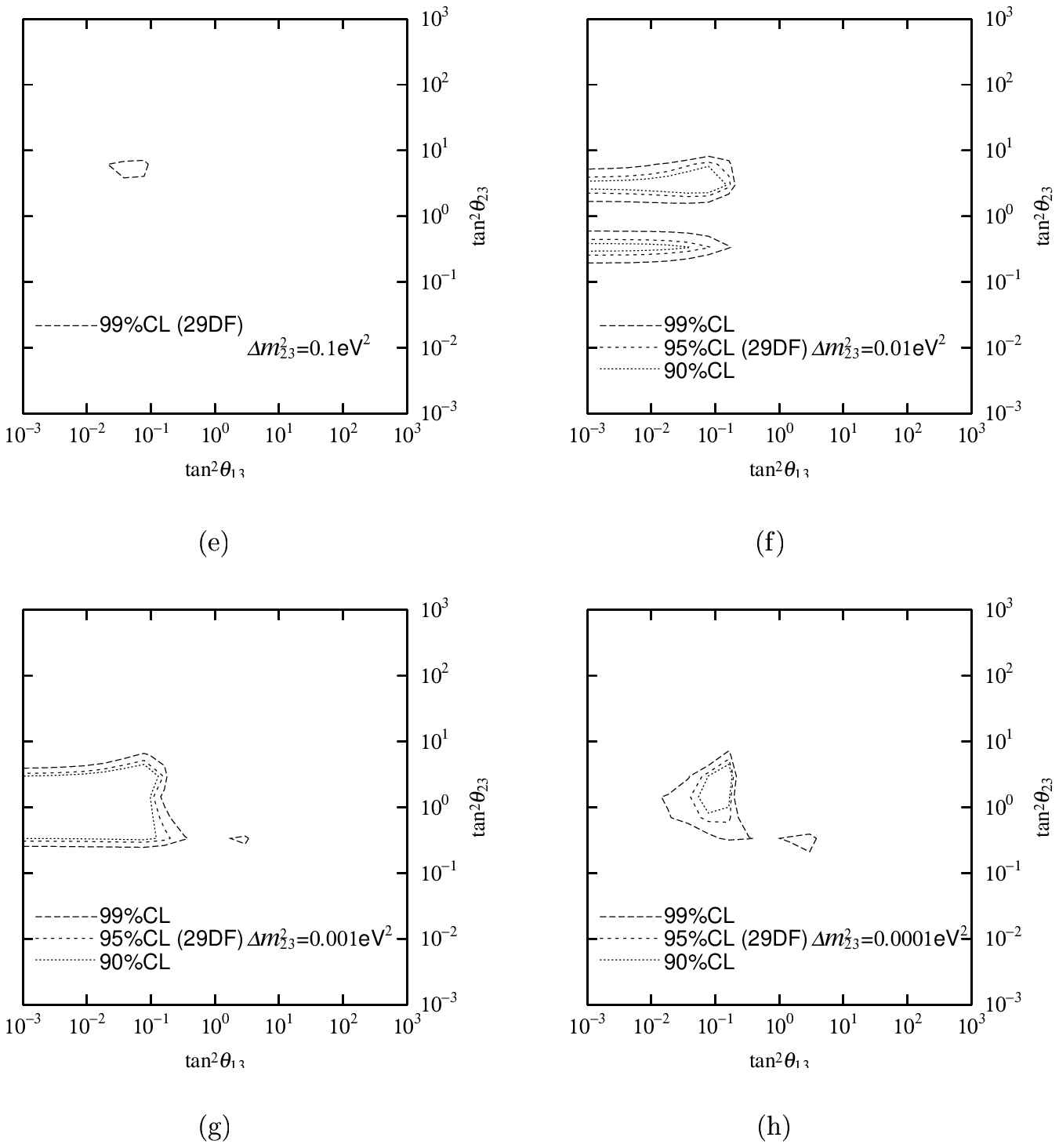}}
%  \figurebox{13cm}{13cm}
   \caption{The plots of allowed regions in the $\tan^2\theta_{13}$--$\tan^2\theta_{23}$ plane for the combinations of sub-GeV, multi-GeV and upward through-going $\mu$ experiments.}
\label{fig3}
\end{figure}  
%%%%%%%%%%%%%%%%%%%%%%%%%%%%%%%%%%%%%%%%%%%%%%%

\subsection{Terrestrial neutrinos}
\vspace{0.2cm}
\par
By many terrestrial experiments, the strong restrictions are imposed to neutrino mixing parameters and masses. In terrestrial experiments, there are two types of experiments, short baseline and long baseline experiments.\cite{K2K} In this paper, we analyze the short baseline experiments. Accelerator experiments searching for the appearance of $\nu_{\tau}$ in $\nu_{\mu}$ were performed by E531, CHORUS and NOMAD \cite{E531}
\begin{eqnarray}
&&P(\nu_{\mu}\to\nu_{\tau})<2\times10^{-3}\ \ (90\%\ {\rm C.L.}),\label{terrmutau}\\
&&\qquad\qquad L/E\sim0.02.\nonumber
\end{eqnarray}
Accelerator experiments searching for the $\nu_{\mu}\to\nu_{e}$ and $\bar{\nu}_{\mu}\to\bar{\nu}_e$ oscillations were accomplished by  E776, KARMEN and LSND \cite{E776}, 
\begin{subequations}
\label{terrmue}
\begin{eqnarray}
&&P(\nu_{\mu}\to\nu_{e})<1.5\times10^{-3}\ \ (90\%\ {\rm C.L.}),\ \ {\rm E776}\label{terrmuea}\\
&&\qquad\qquad L=1{\rm km},\ \ \ \ E \sim 1{\rm GeV},\nonumber\\
&&P(\bar{\nu}_{\mu}\to\bar{\nu}_{e})<3.1\times10^{-3}\ \ (90\%\ {\rm C.L.}),\ \ {\rm KARMEN}\label{terrmueb}\\
&&\qquad\qquad L=17.5{\rm m},\ \ \ \ E<40{\rm MeV},\nonumber\\
&&P(\bar{\nu}_{\mu}\to\bar{\nu}_{e})=3.4\mbox{\small${+2.0\atop-1.8}$}\pm0.7\times10^{-3},\ \  {\rm LSND}\label{terrmuec}\\
&&\qquad\qquad  L=30{\rm m},\ \ \ \ E\sim36-60{\rm MeV}.\nonumber
\end{eqnarray}
\end{subequations}
Accelerator experiment searching for the disappearance of $\nu_\mu$  was carried out by CDHSW experiment and the nuclear power reactor experiments searching for the disappearance of $\bar{\nu}_e$ were carried out by the BUGEY experiment and CHOOZ experiment.\cite{CDHSW}    
\begin{subequations}
\label{terree}
\begin{eqnarray}
&&\nu_\mu{\rm -disappearance\ experiment,}\ \ {\rm CDHSW}  \\
&&0.26{\rm eV}^2<\Delta m^2<90{\rm eV}^2\ {\rm are\ excluded\ for\ maximal\ mixing},\nonumber  \\
&& \qquad\qquad L=130,\ 885{\rm m},\ \ E\sim 3{\rm GeV}. \nonumber\\
&&1-P(\bar{\nu}_{e}\to\bar{\nu}_{e})<10^{-2}\ \ (90\%\ {\rm C.L.}),\ \ {\rm BUGEY}\label{terreea}\\
&&\qquad\qquad L=15,\ 40,\ 95{\rm m},\ \ \ E\sim1-6{\rm MeV}.\nonumber \\
&&P(\bar{\nu}_{e}\to\bar{\nu}_{e})=0.98\pm0.04\pm0.09,\ \ {\rm CHOOZ}\label{terreeb}\\
&&\qquad\qquad L=1{\rm km},\ \ \ E\sim3{\rm MeV}.\nonumber
\end{eqnarray}
\end{subequations}
\par 
We show the contour plots of allowed regions in the $\tan^2\theta_{13}$--$\tan^2\theta_{23}$ plane determined by the probability $P$ expressed in Eq.~(\ref{trans.prob.}) satisfying above experimental data Eqs.~(\ref{terrmutau}), (\ref{terrmue}) (except LSND data (\ref{terrmuec})) and (\ref{terree}) in Fig.~4. Allowed regions are left- and right-hand sides surrounded by curves. The curves represent the boundary of $90$ \% C.L. of $P$. We fixed the parameters $\Delta m_{12}^2$ and $\theta_{12}$ as $\Delta m_{12}^2=10^{-5}{\rm eV}^2$ and $\sin^22\theta_{12}=0.8$, and the various values of parameter $\Delta m_{23}^2$ between 0.1eV$^2$ and 0.0001 eV$^2$. Although we fix the parameters $\Delta m_{12}^2$ and $\theta_{12}$ as $\Delta m_{12}^2=10^{-5}{\rm eV}^2$ and \ $\sin^22\theta_{12}=0.01$, the results are not changed, because the probability $P$ of a terrestrial neutrino is insensitive to the $\nu_e-\nu_\mu$ mixing parameters $\Delta m^2_{12}$ and $\sin^22\theta_{12}$. The dotted lines show the allowed regions restricted by LSND data Eq.\ (\ref{terrmuec}).  The allowed regions determined by LSND data are very restricted, then hereafter we do not consider it.
%%%%%%%%%%%%%%%%%%% fig.4%%%%%%%%%%%%%%%%%%%%%%
\begin{figure}
   \epsfysize=13cm
   \centerline{\epsfbox{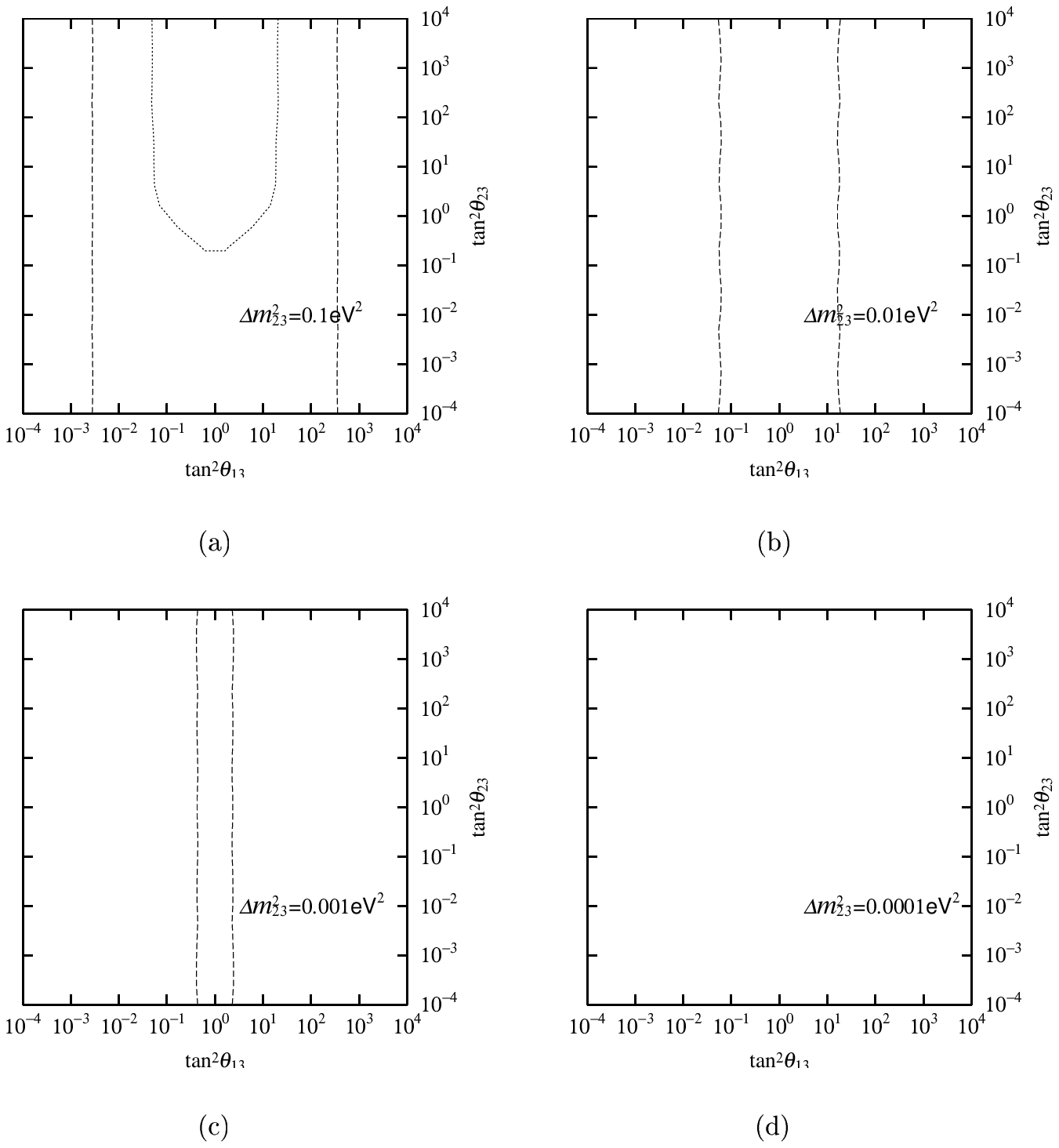}}
%  \figurebox{13cm}{13cm}
   \caption{The plots of allowed regions in the $\tan^2\theta_{13}$--$\tan^2\theta_{23}$ plane determined by $P$ of terrestrial appearance experiments of $\nu_{\mu}\to\nu_{\tau}$, ${\nu}_{\mu}\to\nu_{e}$, $\bar{\nu}_{\mu}\to\bar{\nu}_{e}$  , and the disappearance experiments of $\nu_{\mu}\to\nu_{\mu}$ and $\bar{\nu}_e\to\bar{\nu}_e$. The allowed regions are left- and right-hand sides surrounded by curves. The curves represent the boundary of $90$ \% C.L. of $P$. $\Delta m_{12}^2$ and $\sin^22\theta_{12}$ are fixed as $10^{-5}{\rm eV}^2$ and 0.8, respectively. $\Delta m_{23}^2$ is fixed to 0.1eV$^2$((a)), 0.01eV$^2$((b)), 0.001eV$^2$((c)) and 0.0001eV$^2$((d)). Dotted lines show the allowed regions determined by LSND data.}
\end{figure}  
%%%%%%%%%%%%%%%%%%%%%%%%%%%%%%%%%%%%%%%%%%%%%%%

\subsection{Allowed regions of mixing parameters}
Here, We discuss the allowed regions for mixing parameters $\Delta m^2_{12},\ \sin^22\theta_{12},$ $\Delta m^2_{23},\ \theta_{13}$ and $\theta_{23}$ satisfying atmospheric neutrino and terrestrial neutrino experiments. $\Delta m^2_{12}$ and $\sin^22\theta_{12}$ have been determined in analyses considering the MSW effects of the solar neutrino experiments:\cite{SAGE,FOGLIS,TESHIMA}
\begin{eqnarray}
&&(\Delta m^2_{12},\ \sin^22\theta_{12})\nonumber\\
&&\quad=\left\{\begin{array}{l}
(4\times10^{-6}-7\times10^{-5}{\rm eV}^2,\ 0.6-0.9),\ \ \ {\rm large\ angle\ solution}\\
(3\times10^{-6}-1.2\times10^{-5}{\rm eV}^2,\ 0.003-0.01).\ \ \ {\rm small\ angle\ solution}
\end{array}\right.\\
&& \qquad\qquad {\rm for}\ \theta_{13}=0^\circ-20^\circ\nonumber
\end{eqnarray}
\par
Observing the allowed regions obtained in terrestrial neutrino experiments (Fig. 4) and the allowed regions obtained in atmospheric neutrino experiments including sub-GeV and multi-GeV data (Fig. 1 and 2), we obtain the allowed regions consistent with these experiments. The allowed regions of $\theta_{13}$ and $\theta_{23}$ for the large $\theta_{12}$ angle solution are as follows:
\begin{equation}
\begin{array}{lc}
{\rm for}\ \Delta m_{23}^2=0.1{\rm eV}^2 & \makebox{no allowed region,} \\
{\rm for}\ \Delta m_{23}^2=0.01{\rm eV}^2 & (\theta_{13}<13^\circ,\ 28^\circ<\theta_{23}<37^\circ,\ 53^\circ<\theta_{23}<62^\circ), \\
{\rm for}\ \Delta m_{23}^2=0.005{\rm eV}^2 & (\theta_{13}<9^\circ,\ 28^\circ<\theta_{23}<62^\circ), \\
{\rm for}\ \Delta m_{23}^2=0.002{\rm eV}^2 & (\theta_{13}<23^\circ,\ 29^\circ<\theta_{23}<61^\circ), \\
{\rm for}\ \Delta m_{23}^2=0.001{\rm eV}^2 & (\theta_{13}<20^\circ,\ 29^\circ<\theta_{23}<61^\circ), \\ 
{\rm for}\ \Delta m_{23}^2=0.0005{\rm eV}^2 & (\theta_{13}<24^\circ,\ 35^\circ<\theta_{23}<55^\circ), \\            
{\rm for}\ \Delta m_{23}^2=0.0002{\rm eV}^2 & (4^\circ<\theta_{13}<22^\circ,\ 38^\circ<\theta_{23}<54^\circ),\\
{\rm for}\ \Delta m_{23}^2=0.0001{\rm eV}^2 & \makebox{no allowed region.}  
\end{array}
\end{equation}
For the small $\theta_{12}$ angle solution, similar solution is obtained.
We cannot see the significant difference between the large $\theta_{12}$ angle solution and small one from these results. 
\par
If we adopt the 90\% C.L. of $\chi^2$ for the zenith angle dependence of atmospheric neutrinos, the range of mass parameter $\Delta m^2_{23}$ is restricted to as follows; 
\begin{subequations}
\begin{equation}
\Delta m^2_{23} = 0.01{\rm eV}^2 \sim 0.0002{\rm eV}^2. 
\end{equation}
The value of $\Delta m^2_{23}$ at the minimum of $\chi^2$ under the constraint of $\theta_{13}=0$ is obtained near $\tan^2\theta_{23}=0$ as
\begin{equation} 
\Delta m^2_{23} = 2\times10^{-3}{\rm eV}^2\ \ {\rm with}\ \chi^2=17.
\end{equation}
Under the no constraint of $\theta_{13}$, the value of $\Delta^2_{23}$ at the minimum of $\chi^2$ is obtained at $\tan^2\theta_{13}=6\times10^{-2}$ and $\tan^2\theta_{23}=2$ as
\begin{equation} 
\Delta m^2_{23} = 7\times10^{-4}{\rm eV}^2\ \ {\rm with}\ \chi^2=14.5.
\end{equation} 
\end{subequations}
These results (17a) and (17b) are the same as those of SuperKamiokande (Ref.~\citen{SUPERKAMIOKANDE} and similar to those of a three-flavor neutrinos analysis (the last paper in Ref.~\citen{FOGLIA}). 
\par
The zenith angle distribution of upward through-going $\mu$ flux can be explained by the mixing parameter ranges (16) restricted by the zenith angle distributions of sub-GeV and multi-GeV experiments and terrestrial experiments except for the values near $45^\circ$ of $\theta_{23}$ in large value of $\Delta m^2_{23}$ $(\sim10^{-1}-10^{-2}{\rm eV^2})$. This exception is because the deficit from 1 of the ratio $N_{\rm Exp}(\mu)/N_{\rm MC}(\mu)$ is not so large (about 0.2) in upward through-going $\mu$ flux compared with multi-GeV events. 
\par 
We comment on the present results and ones obtained in our previous analysis.\cite{TESHIMA} In previous study, we analyzed the double ratios of the zenith angle distributions of atmospheric neutrinos neglecting the matter effects of the Earth and the smearing effects, and obtained the results that the mass of $\Delta m^2_{23}$ is rather large and the large $\theta_{12}$ angle solution is favored than the small one. Analyzing the event ratio in present study, the upper limit of the mass parameter $\Delta m^2_{23}$ is limited to low value.  The difference between large $\theta_{12}$ angle solution and small one is disappeared by containing the matter effects of the Earth. 

%%%%%%%%%%%%%% section 4 %%%%%%%%%%%%%%%%%%%%%%%%
\section{Conclusion}
We analyzed the atmospheric neutrino experimental data of SuperKamiokande\cite{SUPERKAMIOKANDE} under the three-flavor neutrino framework with the mass hierarchy $m_1\approx m_2\ll m_3$ and obtained the allowed regions of the parameters $\Delta m_{12}^2,\ \sin^22\theta_{12},\ \Delta m_{23}^2,\ \theta_{13}$ and $\theta_{23}$. We studied the event ratios of the sub-GeV, multi-GeV and upward through-going muons zenith angle distributions. From these atmospheric experiments and terrestrial ones,  the allowed regions of mass parameters $\Delta m^2_{23}$ are restricted as  0.01{eV$^2$}-0.0002{eV$^2$}, and the value of $\Delta m^2_{23}$ at the minimum $\chi^2$ under the restriction $\theta_{13}=0$ is $2\times10^{-3}{\rm eV^2}$. For mixing parameters, there is no difference between the large $\theta_{12}$ angle solution and the small $\theta_{12}$ one. Obtained results are $\theta_{13}<13^\circ,\ 28^\circ<\theta_{23}<37^\circ$ and $53^\circ<\theta_{23}<62^\circ$ for $\Delta m_{23}^2=0.01{\rm eV}^2$, $\theta_{13}<23^\circ$ and $29^\circ<\theta_{23}<61^\circ$ for $\Delta m_{23}^2=0.002{\rm eV}^2$, $\theta_{13}<24^\circ$ and $35^\circ<\theta_{23}<55^\circ$ for $\Delta m_{23}^2=0.0005{\rm eV}^2$, $4^\circ<\theta_{13}<22^\circ$ and $38^\circ<\theta_{23}<54^\circ$ for $\Delta m_{23}^2=0.0002{\rm eV}^2$. In our three neutrino analysis for the SuperKamiokande atmospheric neutrino experiments, the large $\nu_\mu-\nu_\tau$ mixing is favored  and the small mixing $\nu_e-\nu_\mu$ mixing is allowed at the same time. This result will be significant in the long baseline experiments.

%%%%%%%%% Appendix %%%%%%%%%%%%%%%%%%%%%%%%%%%%
\appendix
\section{Explicit calculation of Eqs.(10a) and (10b)}
 In multi-GeV case, the information of the Super-Kamiokande detection efficiency $\epsilon_\alpha(E_\alpha)$ is not known for us, but the detected charged lepton event number $f_\alpha(E_\alpha)$ is given in the paper of SuperKamiokande collaboration.\cite{FUKUDA} The event number $f_\alpha(E_\alpha)$ is expressed as    
\begin{equation}
f_\alpha(E_\alpha)=\int\epsilon_\alpha(E_\alpha)\sigma_\alpha(E_\nu,E_\alpha,\psi)F_\alpha(E_\nu,\theta-\psi)dE_\nu d\cos\psi d\cos\theta. 
\end{equation}
We define the effective zenith angle dependent neutrino flux $f_\alpha(E_\alpha,\theta)$ as
\begin{equation}
f_\alpha(E_\alpha,\theta)=f_\alpha(E_\alpha)\frac{F_\alpha(E_\alpha,\theta)}{\int^1_{-1}d\cos\theta F_\alpha(E_\alpha,\theta)},
\end{equation}
where we use the zenith angle distribution given in Ref.~\citen{HONDA} for $F_\alpha(E_\alpha,\theta)$. Furthermore we approximate the $\psi$ dependence of differential cross section $\sigma_\alpha$ by a smearing Gaussian function $n(\psi)$ as
\begin{equation}
n(\psi)=\frac{2}{\sqrt{\pi}\tan(\psi_0/2)}\frac{1}{(1+\cos\psi)\sin\psi}\exp\left\{-\frac{\tan^2(\psi/2)}{\tan^2(\psi_0/2)}\right\},
\end{equation}
where the resolution $\psi_0$ is taken as $17^\circ$.  
Then, we estimate ${dN_{\rm Exp}}/{d\cos\theta}$ by the following integral, 
\begin{equation} 
\frac{dN_{\rm Exp}}{d\cos\theta}=\int n(\psi)f_\beta(E_\beta,\theta-\psi)P^M(\nu_\beta\to\nu_\alpha)dE_\beta d\cos\psi.
\end{equation}
\par
In sub-GeV case, $\sigma_\alpha(E_\nu)=\int\sigma_\alpha(E_\nu,E_\alpha,\psi)d\cos\psi dE_\alpha$ is calculated in Ref.~\citen{KAJITA}. $\epsilon_\alpha(E_\alpha)$ is nearly 1 in the sub-GeV region, thus we estimate the detected lepton event number with zenith angle dependence $f_\alpha(E_\nu,\theta)$, approximating $\theta=\psi$ in $F_\alpha(E_\nu,\theta-\psi)$,
\begin{equation}
f_\alpha(E_\nu, \theta)=\epsilon_\alpha(E_\nu)\int\sigma_\alpha(E_\nu,E_\alpha,\psi)F_\alpha(E_\nu,\theta)dE_\alpha d\cos\psi=\epsilon_\alpha(E_\nu)\sigma_\alpha(E_\nu)F_\alpha(E_\nu,\theta),
\end{equation} 
where information of $F_\alpha(E_\nu,\theta)$ is taken from Ref.~\citen{HONDAP}.  Then, introducing the smearing Gaussian function $n(\psi)$ as the multi-GeV case, we express $dN_{\rm Exp}/d\cos\theta$ as
\begin{equation} 
\frac{dN_{\rm Exp}}{d\cos\theta}=\int n(\psi)f_\beta(E_\nu,\theta-\psi)P^M(\nu_\beta\to\nu_\alpha)dE_\nu d\cos\psi,
\end{equation}
where the resolution $\psi_0$ is taken as $60^\circ$ for sub-GeV case. 
\par 
For upward through-going muons fluxes, event number $f_\alpha(E_\alpha)$ of through-going muons are estimated in Ref.~\citen{KAJITA}. Thus, using this information and the same treatment as multi-GeV case, we calculate the zenith angle $\theta$ dependent events of upward through-going muons.

%%%%%%%% references %%%%%%%%%%%%%%%%%%%%%%%%%

\end{document}